
\documentstyle{amsppt}

\pagewidth{14cm}
\pageheight{22cm}
\hcorrection{9mm}
\vcorrection{8mm}
\refstyle{C}

\rightline{DAMTP R92-46}
\rightline{CTP TAMU-7/93}
\rightline{NOTT-MATHS 93/1}
hep-th/9302073
{}
\bigskip
\topmatter
\title Kleinian Geometry and \\ the N=2 Superstring.\endtitle
\author J. Barrett$^{\ast}$, G.W. Gibbons$^{\star}$, M.J.
Perry$^{\star}$, C.N.Pope$^{\dag}$ and
P.Ruback$^{\ddag}$ \endauthor
\affil
{}$^{\ast}$ Department of Mathematics \\ University of Nottingham \\
University Park\\ Nottingham \\ NG7 2RD \\ England \\ \\ \\
{}$^{\star}$ Department of Applied Mathematics and Theoretical
Physics\\ University of Cambridge \\
Silver Street \\ Cambridge \\ CB3 9EW \\England \\ \\ \\
{}$^{\dag}$Department of Physics \\ Texas A and M University \\ College
Station \\ Texas 77843 \\ USA \\ \\ \\
{}$^{\ddag}$ Airports Policy Division\\Department of  Transport\\
2 Marsham Street\\
London \\ SW1P 3EB\\ England \\
\endaffil
\abstract
This paper is devoted to the exploration of some of the geometrical
issues raised by the $N=2$ superstring. We begin by reviewing the reasons
that $\beta$-functions for the $N=2$ superstring require it to live in a
four-dimensional self-dual spacetime of signature $(--++)$, together with
some of the arguments as to why the only degree of freedom in the theory
is that described by the gravitational field. We then move on to describe
at length the geometry of flat space, and how a real version of twistor
theory is relevant to it. We then describe some of the more
complicated spacetimes that satisfy the $\beta$-function equations.
Finally we speculate on the deeper significance of some of these
spacetimes.
\endabstract
\endtopmatter
\vfill\eject
\document

{\bf \S 1 Introduction}
\bigskip

This paper is aimed towards providing a description of some of
the physics of the $N=2$ superstring, together with some of the
geometrical consequences of such theories. It  follows on from
some seminal work of Ooguri and Vafa \cite{1} who showed how to make
sense of this type of string theory. They showed that the critical
dimension of such strings is four real dimensions, and then computed some
scattering amplitudes. The amplitudes indicate that the bosonic part
of the $N=2$ theory corresponds to self-dual metrics of a spacetime of
ultrahyperbolic signature $(++--)$. We call metrics of such signature
Kleinian. The situation should be contrasted with the bosonic string
where we expect 26-dimensional metrics of Lorentz signature to play an
important role, and the $N=1$ superstring where 10-dimensional metrics
of Lorentz signature are the important objects. If we try to extend
the number of supersymmetries beyond two superconformal supersymmetries,
then we find that the critical dimensions are negative. It is unclear
how one should think about such types of string theory.

The present paper is divided into six sections. In
section two, we discuss some of the basic properties of the $N=2$
superstring, and explain how self-dual spaces of dimension four arise.
In section three, we discuss the properties of flat four-dimensional
Kleinian spacetime, and see how a real version of
twistor theory is relevant. In section four we discuss some curved
metrics that arise in this type of geometry. In section five, we discuss
some cosmic string metrics which are in some sense related to the planar
strings, and in section six we discuss some metrics that are related to
string of higher genus. We conclude with some
speculations. Readers who are more interested in Kleinian geometry than
superstrings may skip section two and proceed directly to section three.
\bigskip

{\bf \S 2 The N=2 String}
\bigskip

The traditional view of string theory is to regard the string worldsheet
as an object embedded in a classical spacetime background. If we have no
other structure then we would be thinking about the bosonic string. In
perturbation theory, the closed bosonic string describes gravity with
the Einstein action, together with various other spacetime fields.
Such a theory is both conformally and Lorentz invariant in
26-dimensional flat spacetime of Lorentz signature. The bosonic string
however is believed to be inconsistent because its spectrum contains a
tachyon \cite{2}.

A different string theory can be found by supposing that the worldsheet
has some additional structure. If it has $N=1$ superconformal
supersymmetry, we find that the string is quantum mechanically
consistent in a 10-dimensional spacetime of Lorentz signature. The closed
superstring describes gravity with the Einstein action coupled to
various other spacetime fields, the massless  fields being those found
in either of the two $N=2$ supergravity theories. There is no longer a
tachyon in its spectrum and it is believed that this theory is a
self-consistent finite theory. Nevertheless the description in terms
of the worldsheet is incomplete  because, like the
bosonic  string, the density of states function $\rho(E)dE$, the number
of physical spacetime states between $E$ and $E+dE$, grows
exponentially. As a consequence, there is a maximum temperature, the
Hagedorn temperature, at which there is a phase transition to some other
\lq\lq exotic" phase of string theory \cite{3}. For the superstring this
temperature is $T_H \sim (8\pi^2\alpha')^{-1}$ where $\alpha'$ is the
inverse string tension. The nature of the new phase is not currently
understood - although much has been made of the analogy between hadronic
physics and QCD. It may be extremely difficult to resolve the issue
problem for the superstring precisely because of the complexity of the
theory.

What is needed is a much simpler model theory in which some of the issues
can be investigated without the complications of the infinite number of
degrees of freedom of the $N=1$ string. The model we propose to
investigate here is the $N=2$ superstring, which is in some sense a
\lq\lq topological" string theory. The $N=2$ superstring is a theory
with two superconformal supersymmetries. The action for such a theory
can be conveniently written so that the two supersymmetries can be
combined into a single complex supersymmetry. If we let $Z^i$ be the
$i$-th component of a complex bosonic field and $\psi^i$ be the $i$-th
component of a complex world-sheet spinor field (i.e. a Dirac spinor)
then the action
$$ I= {1\over 2\pi\alpha'}\int \gamma^{1/2} d^2\sigma
\{\gamma^{\mu\nu}\partial_{\mu}\bar Z^i \partial_{\nu}Z^j - i
\bar\psi^i\gamma^{\mu}\partial_{\mu}\psi^j \}\eta_{ij} \eqno(2.1) $$
is invariant under an $N=2$ rigid supersymmetry.
In this expression $\alpha'$ is the inverse string tension, and the
string world-sheet is spanned by coordinates
$\sigma^{\mu}$, and has metric tensor $\gamma_{\mu\nu}$. Gamma matrices
are determined with respect to the metric $\gamma_{\mu\nu}$. The
spacetime metric is
given by $\eta_{ij}$ and has not, as yet, been specified. The rigid
supersymmetry
is generated by a complex (Dirac) spinor $\epsilon$, and
the transformations of $Z^i$ and $\psi^i$ are given by
$$ \delta Z^i = \bar\epsilon\psi^i, \eqno(2.2)$$
$$ \delta\psi^i = -i(\partial_\mu Z^i)\gamma^\mu\epsilon. \eqno(2.3)$$
The global supersymmetry can be promoted to a local supersymmetry
by coupling the matter multiplet to $N=2$ world-sheet supergravity. The
graviton multiplet contains a complex gravitino $\chi_\mu$ and an $U(1)$
gauge field $A_\mu$ besides
the world-sheet metric $\gamma_{\mu\nu}$. The action is now invariant
under world-sheet reparametrizations and world-sheet Weyl rescalings, and
can be written explicitly as \cite{4} $$\eqalign{ I=& -{1\over 2\pi
\alpha'} \int \gamma^{1\over 2} d^2\sigma \{
\gamma^{\mu\nu}\partial_{\mu}\bar Z^i\partial_{\nu}Z^j
-i\bar\psi^i\gamma^\mu\partial_\mu\psi^j
             +i(\partial_\mu\bar\psi^i)\gamma^\mu\psi^j\cr
& + A_\mu \bar\psi^i \gamma^\mu \psi^j
+  (\partial_\mu Z^i - {1\over
2}\bar\chi_\mu\psi^i)\bar\psi^j\gamma^\nu\gamma^\mu\chi_\nu \cr
&+ (\partial_\mu\bar Z^i - {1\over
2}\bar\psi^i\chi_\mu)\bar\chi^\nu\gamma^\mu\gamma^\nu\psi^j \}
\eta_{ij}. \cr} \eqno(2.4)$$
There are no kinetic terms for any components of the N=2 supergravity
multiplet, and in fact these fields can be gauged away by the
introduction of the appropriate ghost fields. As it stands the action
(2.4) describes a string propagating in a complex ${1\over
2}d$-dimensional background. However, it can be reinterpreted as a
$d$-dimensional real background, where the signature
of the metric $\eta_{ij}$ has an even number of both  positive and
negative directions so as to be consistent with the complex structure
inherent with $N=2$ supersymmetry.

We must now examine the consistency of our string theory. Suppose that
the string is propagating in a flat real metric $\eta_{ij}$. First, there
is the issue of the Weyl anomaly. It is easily computed, since the
contribution to the anomaly from a ghost-antighost pair in which
the ghost field has conformal weight ${1\over 2}(1-j)$ and statistics
$\epsilon,$ is $ \epsilon(1-3j^2)$. The graviton ghost has $\epsilon=+1$
and $j=3$ thus contributes  $-26$ to the anomaly. There are two sets of
gravitino ghosts, one for each gravitino both with $\epsilon=-1$ and
$j=2$ giving a contribution of $22$. The ghost for the $U(1)$ gauge
field has $\epsilon=1$ and $j=1$ and so contributes $-2$. The total
ghost contributions are therefore $-6$. The matter fields contribute
$d$ from the bosonic fields and ${d/2}$ from the fermionic fields
and thus the Weyl anomaly will cancel provided that $d=4$. Thus we
conclude that the $N=2$ string is anomaly-free in a flat spacetime of
four dimensions. The choice of metric signature is however not
completely arbitrary. The target space must have a complex structure and
so we can choose the target space metric to have signature 4 or 0.
However, if the worldsheet is to be Lorentzian, then
the metric signature is $(--++)$. We will briefly comment about
the Euclidean signature case later.

The next step is to determine the physical states of the theory. These
are the states of positive norm modulo gauge equivalence. For signature
$(--++)$ one expects twice as many negative norm states as for the
bosonic string, and all of these must be removed by the various ghost
contributions. In fact, there are two sets of fermionic ghosts, those
that cancel the graviton, and those that cancel the gauge field. These
are sufficient to cancel the negative norm states. On the face of it,
{\it all} the states of the theory are cancelled, since each ghost field
absorbs both a timelike and a spacelike degree of freedom. It is in this
sense that we mean the $N=2$ superstring is a topological string theory.

We must however be more precise, and look at the question of degrees of
freedom more carefully. There are two equivalent ways of proceeding.
One is to examine the cohomology of the BRS operator defined on a
cylindrical worldsheet. Physical states are then those that
are $Q$-closed but not $Q$-exact \cite{5}. Alternatively one can look at
the partition function evaluated on the torus. This will exhibit
contributions from each physical state. Unitarity considerations
indicate that the two approaches yield the same conclusions. In going
through such a process, it is clear that there is considerable ambiguity
in how one deals with the global issues, and it appears that there are
in fact many consistent string theories that might eventually emerge.

To see what the physical states are in the BRS formulation, we must
introduce sets of operators with which to describe the excited states
of the string. We are interested in closed strings propagating in the
flat metric $\eta_{ij}$ on $\Bbb R^4$. The string worldsheet is
cylindrical with spacelike coordinate $\sigma$ identified with period
$2\pi$, and the timelike coordinate $\tau$ being unbounded. We work,
for convenience, with complex fields, $X^i$ and $\psi^i$. It therefore
follows that $X^i$ is periodic
$$ X^i(0,\tau) = X^i(2\pi,\tau), \eqno(2.5)$$
and $\psi^i$ is either periodic or antiperiodic
$$ \psi^i(0,\tau) = \pm \psi^i(2\pi,\tau). \eqno(2.6) $$
 Introducing mode expansions for
these fields
$$ X^i = x_0^i + p^i\tau + i \sum_{n\ne 0}{1\over n}(\alpha_n^i
e^{-in(\tau+\sigma)} + \tilde\alpha_n^i e^{-in(\tau-\sigma)}),
\eqno(2.7)$$
and
$$ \psi^i = \sum_r d^i_r e^{-in(\tau+\sigma)} +
            \tilde d^i_r e^{-in(\tau-\sigma)}, \eqno(2.8) $$
where $r \in \Bbb Z$ (R sector) or $r \in \Bbb Z +{1/2}$ (NS sector).
The canonical  (anti)-commutation rules then lead to
$$ [\alpha_n^i,\bar\alpha_m^j] = n\eta^{ij}\delta_{n+m,0} \eqno(2.9)$$
and
$$ \{d_r^i,\bar d_s^j\} = \eta^{ij} \delta_{r+s,0}, \eqno(2.10)$$
together with similar expressions for the left-handed modes.
All other (anti)-commutators vanish.

As a result of $N=2$ superconformal symmetry the energy-momentum tensor
$T_{\mu\nu}$, the (complex) supercurrent, $S_\mu$ and the $U(1)$
gauge current $T_\mu$ must all vanish identically.
$$ T_{\mu\nu} = [\partial_\mu X^i \partial_\nu X^j +
\bar\psi^i\gamma_{(\mu}\partial_{\nu)}\psi^j - (trace)\ ] \eta_{ij}.
\eqno(2.11)$$
$$ S_\mu = \gamma^\nu\gamma_\mu\bar\psi^i\partial_\nu X^j \eta_{ij}.
\eqno(2.12) $$
$$ T_\mu = \bar\psi^i\gamma_\mu\psi^j\eta_{ij}. \eqno(2.13)$$
The Fourier components of these are given by the following operator
expressions, each of which should be taken to be normal ordered in the
quantum theory
$$ L_n = \sum_m \alpha_m\bar\alpha_{n+m} + \sum_r(r+{n\over
2})d_{-r}\bar d_{n+r}, \eqno(2.14)$$
$$ G_r = \sum_s d_s\alpha_{r-s}, \eqno(2.15)$$
$$ \bar G_r = \sum_s \bar d_s \bar\alpha_{r-s}, \eqno(2.16) $$
$$ T_n = \sum_r d_r \bar d_{n-r}. \eqno(2.17) $$

The $N=2$ superconformal algebra can then be constructed in the
standard way. It is, taking care of all anomalous (central) terms
$$ [L_n,L_m] = (n-m) L_{n+m} + {1\over 2}n(n^2-1)\delta_{n+m,0},
\eqno(2.18)$$ $$ [L_n,T_m] = -m T_{n+m}, \eqno(2.19) $$
$$ [L_n,G_r] = ({1\over 2}n-r)G_{n+r}, \eqno(2.20)$$
$$ [L_n,\bar G_r] = ({1\over 2}n-r)\bar G_{n+r}, \eqno(2.21)$$
$$ [T_n,G_r] = G_{n+r}, \eqno(2.22)$$
$$ [T_n,\bar G_r] = -\bar G_{n+r}, \eqno(2.23)$$
$$ [T_n,T_m] = 2m\delta_{n+m,0}, \eqno(2.24)$$
$$ \{G_r,\bar G_s \} = L_{r+s} + {1/2}(r-s)T_{r+s} +
(r^2-{1/4})\delta_{r+s,0},  \eqno(2.25)$$
the remaining (anti)-commutators all vanish.

The physical states conditions are then
\def\stp{\mid\phi\rangle}
$$ L_n\stp = T_n\stp = G_r\stp =\bar G_r\stp = 0 ~~~n,r >0, \eqno(2.26)$$
$$ T_0 \stp =0, \eqno(2.27) $$ $$ L_0\stp = h\stp, \eqno(2.28)$$
for a physical state $\stp$. The conditions (2.26-8) are precisely
equivalent to the requirement that $\stp$ be described by the cohomology
of the BRS operator $Q$. It should be noted that these conditions imply,
from equation (2.25), that all physical states have vanishing $U(1)$
charge, whereas the currents $G$ and $\bar G$ have a $U(1)$ charge of
$1$ and $-1$ respectively.

The highest weight representations of the $N=2$ super-Virasoro algebra
have been investigated by Boucher, Friedan and Kent \cite{6}. We can
apply their results directly to the string problem. For the NS-sector,
the only possible choice of $h$ consistent with the requirement of
neutrality with respect to the $U(1)$ charge {\it for the ground state},
and central charge corresponding to the physical dimension in which
conformal invariance can be maintained is $h=0$. Furthermore, beyond the
ground state, all excited states formed by acting with the $\alpha_n$
\rq s or $\bar\alpha_n$ \rq s on the vacuum state will not be neutral
with respect to the $U(1)$ charge, or spurious states. The conclusion is
therefore that the only state that represents a spacetime boson is the
ground state of the string, which is a massless.
This agrees with the results of Mathur and Mukhi \cite{7} who examined
the first few excited states of this superstring.

In the R-sector of the string, their results can be similarly applied,
and we find no physical states whatsoever. Thus we conclude that the
only physical state of the string is a single massless scalar boson.
This string therefore does not suffer from a Hagedorn transition
together with its associated complications.

If instead we wished to derive the same results from considering the
cohomology of $Q$, then we would have to supplement the definitions of
$L_n, G_n, \bar G_n $ and $T_n$ by the inclusion of extra ghost terms.
We would then discover that the (2.18) to (2.25)
would still hold, but now all the central terms would cancel, in both
the NS and R sectors, as a consequence of choosing the critical
 dimension to be four, and of having zero shift of the vacuum
energy in both sectors. We  note that the above situation is rather
different to  the $N=1$ superstring where there are
different shifts of the vacuum energies in different sectors. Having
carried out this procedure, we would be guaranteed that $Q^2=0$, and
then its cohomology could be investigated. Thus the
analysis of the previous paragraph would be repeated.

The interpretation of this scalar particle has been explored by Vafa
and Ooguri \cite{1} who have computed the scattering amplitudes
at the tree level. They find that  the amplitudes
obtained show that the scalar is the K\"ahler potential for a
self-dual gravitational field. Thus it seems reasonable to conclude
that this string theory is just self-dual gravity,
for the case of spacetime signature $(--++)$.

One can also support their conclusion by two other calculations.
Firstly, we can compute the string partition function
directly from the path integral which requires the explicit introduction
of the ghost and antighost fields for each of the constraints,
given by equations (2.14-17). The ghosts are just minimal b-c systems
with statistics $\epsilon$ and conformal weight $j$ referred to earlier.
We can now simply evaluate the partition function by standard methods.
Consider states at level $n$ so that the $mass^2$ of these states is
proportional to $n$, and define the partition function
to be
$$ G(w) = Tr~ w^N = \sum p(n) w^n \eqno(2.29)$$
where $N$ is the number operator which counts the level to which the
string has been excited, and thus $p(n)$ is the number of states at
the $N^{th}$ level. Since we are dealing with a closed string theory, we
must take account of both left-moving excitations and right-moving
excitations, imposing the constraint that the level of excitation of
the left-movers is the same as that of the right-movers.
Consider first of all the NS-sector (for just right-movers),
then
$$ G_{NS}(w) = \prod_{n>0} (1-w^n)^{-4} (1+w^{n+{1/2}})^4 (1+w^n)^4
(1+w^{n+{1/2}})^{-4} \eqno(2.30)$$
where each term in the product arises from the $X^i$'s,
$\psi^i$'s, graviton and gauge field ghosts, and gravitino ghosts
respectively. We thus find that $G_{NS}(w)=1$ as was expected from
previous considerations.
Now we shall consider the R-sector. There are no  states that
have vanishing charge, and so we can conclude immediately that
$G_R(w)=0$. Hence, we believe the only physical state is the massless
scalar. The fact that $G(w)=1$ is the reason that the theory can be
termed  topological.

Of course, we have chosen one particular GSO projection \cite{8} in
the preceeding treatment,  and it is the one in which all the fermions
have the same choice of periodicity. We need not have done that, and
Ooguri and Vafa \cite{1} exhibit other possible choices. We note here
however that if we make a choice that results in spacetime fermions,
then there will be at least two spacetime supersymmetries that are never
broken. That is because in a self-dual space, we are guaranteed to have
at least two covariantly constant spinors which can be used to generate
two independent spacetime supersymmetries.

We have so far examined the theory in flat space only. Suppose that
instead we started from the $\sigma$-model type action and asked what
would happen if the model were immersed in some curved space. Then, the
would need to compute the renormalization group equations for the
curved space metric $\eta_{ij}$. At the one-loop level, we would find
that the $\beta$-function equation for the renormalization of the metric
would just be
$$ R_{ij}=0 \eqno(2.31)$$
Thus to lowest order, we have discovered that the
metric must be Ricci flat. Stringy considerations lead us to believe
that this metric must be K\"ahler because of the behaviour of the
scattering amplitudes alluded to earlier. Thus, in perturbation theory
we can conclude that there are no other terms. Suppose that we have a
Ricci flat K\"ahler metric, since we are four dimensions we can conclude
that the metric must be self-dual. It is therefore impossible to find any
non-trivial higher order symmetric tensors built from only the Riemann
tensor and so any such target space will solve the string
$\beta$-function equations to all orders. Consequently in most of the
remainder of our paper, we will describe some of the geometry of such
target spaces.

We conclude this section by contemplating the string propagating in a
space of signature $(++++)$ once more. We have discovered that the only
physical excitations of the string were the ground state. If we were in
flat $\Bbb R^4$ then with this signature, there would be no string
theory as the ground state would always correspond to the string having
collapsed down to a point. However, if the target space, $\Cal M$ had
some non-trivial $\pi^2$, then the string at genus zero could shrink down
onto the non-contractible $S^2$ and give some extra contribution to the
partition function. We can extrapolate  our present results to conclude
that
$$ G_{NS}(0) = dim(\pi^2(\Cal M)) \eqno(2.32) $$
in general. Furthermore, similar considerations at genus $g$ would
seem to indicate that all of $H_2(\Cal M)$ can be probed by considering
all possible genera of the string worldsheet.
It may even be possible to extract further topological
information from $N=2$  string theory. This is a matter that is
currently under investigation.

\bigskip

\def\we{{}_\wedge}
{\bf \S 3 Flat Ultra-hyperbolic 4-Geometry}
\bigskip
The properties of four-dimensional metrics of signature (2,2) are
comparatively unknown amongst most of the physics community. The purpose
of this section therefore is to provide a brief review of their basic
properties. We shall begin with the flat space $\Bbb R^{2,2}$, and
discuss curved spaces in the next section.

We will start with some purely terminological remarks. In an arbitrary
dimension, an ultrahyperbolic metric is a metric with signature $(s,t)$
or $(t,s)$ with $s \ge t$ and for which $t \ne 0,1$. The cases of
$t=0,1$ are called Riemannian and
Lorentzian (or hyperbolic) metrics respectively. In dimension four,
there remains only the case $(2,2)$. In dimensions greater than four, the
term  \lq\lq ultrahyperbolic," as it is usually used, is ambiguous.
Another case that is especially interesting for us is that of $(3,3)$,
since $GL(4,\Bbb R)$ is a 2-fold cover of $SO(3,3;\Bbb R)$ which itself
is a 2-fold cover of the conformal group of $\Bbb R^{2,2}$. We propose
calling metrics of signature $(r,r)$ \lq\lq Kleinian." We shall also
refer to a flat space of signature $(r,r)$ as \lq\lq Pl\"ucker" space by
analogy with flat Euclidean space and flat Minkowski spacetime. Not the
least interesting feature of Kleinian spacetimes is the complete
symmetry between space and time which corresponds to the freedom to
multiply the metric $g_{\mu\nu}$ by $-1$ and obtain a new metric of the
same signature. Only in dimension 2 is it possible for a Lorentzian
spacetime because in that dimension Lorentzian and Kleinian are the same
thing. In string theory the above symmetry is associated with what is
usually referred to as  \lq\lq crossing symmetry." Since the analogy is not
precise and since the word \lq\lq dual" is used in many unrelated situations,
we prefer to use the term \lq\lq chronal-chiral symmetry" for the symmetry
under $g_{\mu\nu} \rightarrow -g_{\mu\nu}.$

The reason for the introduction of the names of Klein \cite{9} and
Pl\"ucker \cite{10} is of course because of the well-known Pl\"ucker
correspondence between unoriented lines in Euclidean 3-space $\Bbb R^3$
and null rays through the origin in $\Bbb R^{3,3}$, via the
corresponding relation to simple 2-forms in $\Bbb R^{4-t,t}$ for
all values of $t$, and its
exploitation and popularization by Klein. The Pl\"ucker correspondence
has numerous important applications to the geometry and physics of
Euclidean 3-space. The extension of the Pl\"ucker correspondence to
4-dimensional Minkowski space is what Penrose \cite{11} calls twistor
theory \cite{12}. The Pl\"ucker correspondence works as follows: Let
$X^\alpha, Y^\beta$ be homogeneous coordinates for two points $x$ and
$y$ in $\Bbb R^3$, or more precisely $P_3(\Bbb R)$. The line $l$ through
$x$ and $y$ corresponds to the simple bi-vector
$$ P^{\alpha\beta} = 2X^{[\alpha}Y^{\beta]},  \eqno(3.1) $$
up to a scale factor. The condition that the bi-vector be simple is
that
$$ \epsilon_{\alpha\beta\gamma\delta}P^{\alpha\beta} P^{\gamma\delta}=0,
\eqno(3.2)$$  where the left hand side of (3.2) defines a Kleinian metric
on the six-dimensional space of all bivectors in $\Bbb R^4.$ Equation
(3.2)  says that lines in $P_3(\Bbb R)$ correspond to null rays
through the origin in $\Bbb R^{3,3}.$ The six numbers $P^{\alpha\beta}$
subject to the constraint (3.2) are referred to as the Pl\"ucker
coordinates for the straight line $l$.

If $X^{\alpha} = (1, x^i)$ etc. then
$$ P^{i0} = x^i - y^i, \eqno(3.3a)$$
$$ P^{ij} = x^i y^j - x^j y^i. \eqno(3.3b)
$$ Using a suitable numbering system, we may introduce
coordinates  $p^a$ in $\Bbb R^{3,3}$ and equation (3.2) becomes
$$ p^a p^b k_{ab}  =0,\eqno(3.4) $$
where $k_{ab}$ are the components of the Pl\"ucker metric.

Thus if
$$\eqalign {&\epsilon_{0123} = 1,\cr
&p^i = P^{0i} \cr
&p^{3+i} = {1\over 2}\epsilon^{ijk} P^{jk}, \cr} $$
then
$$ k_{ab}= \delta_{3,\,\mid a-b \mid}. \eqno(3.5) $$

The action of the projective group of $\Bbb R^3$, $PSL(4,\Bbb R) \cong
{SL(4;\Bbb R)/ \pm 1} $, may be lifted to a linear action on $\Bbb
R^4$ preserving the alternating tensor
$\epsilon_{\alpha\beta\gamma\delta}$. This provides the isomorphism
$SO_0(3,3;\Bbb R) \cong {SL(4,\Bbb R)/ \pm 1}$, the elements of
$\Bbb R^4 $ corresponding to what are called in the physics literature
Majorana-Weyl spinors for $SO(3,3;\Bbb R).$

The space of lines in $P_3(\Bbb R)$, $Gr_2(\Bbb R^4)$ carries a natural
conformal structure: two lines $l_1$ and $l_2$ are null separated iff
they intersect. Equivalently, their respective
bivectors satisfy
$$ \epsilon_{\alpha\beta\gamma\delta} P_1^{\alpha\beta}
P_2^{\gamma\delta} = 0, \eqno(3.6) $$
that is to say that
$$ p_1^a p_2^b k_{ab}  = 0. \eqno(3.7)$$
In other words the Kleinian 6-metric $k_{ab}$ on $\Bbb R^{3,3}$
induces on the 4-dimensional space of lines $Gr_2(\Bbb R^4)$ this
conformal structure. Topologically $Gr_2(\Bbb R^4)$ is ${(S^2 \times
S^2)/\pm 1}$ where the action of $\pm 1$ is the antipodal map on each
factor. If we set
$$ a_i = p^i + p^{i+3} = P^{0i} + {1\over 2}\epsilon^{ijk}P^{jk},
\eqno(3.8a) $$
$$ b_i = p^i - p^{i+3} = P^{0i} - {1\over 2}\epsilon^{ijk}P^{jk},
\eqno(3.8b)$$
the light cone condition (3.2), or equivalently (3.4), is
$$ {\bold a}^2 = {\bold b}^2, \eqno(3.9)$$
whilst the pairs $({\bold a}, {\bold b})$ and $(\lambda {\bold a}
,\lambda{\bold b}) , \lambda\ne 0$ must be identified. Choosing
$\lambda$ to be positive we may set,
without any loss of generality, ${\bold a}^2={\bold b}^2 = 1 $
and the remaining freedom is $(\bold a,\bold
b)\rightarrow(-\bold a,-\bold b)$.  This metric is conformally
flat, but not Einstein, and shows in fact that we may regard ${S^2
\times S^2/\pm1}$ as the conformal compactification of flat Pl\"ucker
4-spacetime $\Bbb R^{2,2}$, just as ${S^1\times S^3/\pm1}$ and
${S^4/\pm1}$ with their standard product metrics provide the conformal
compactifications of Minkowski spacetime and Euclidean space
respectively. To obtain flat Pl\"ucker 4-spacetime itself, we consider
not all null rays in Pl\"ucker 6-spacetime $\Bbb R^{3,3}$, but those
which intersect a null hypersurface. The rays intersecting a timelike or
a spacelike hyperplane correspond to the conformally flat space with
constant curvature and signature $(2,2)$ whose isometry group is
$SO(3,2;\Bbb R)$, that is to the quadric in $\Bbb R^{3,2}$ given by
$$  (x^1)^2 + (x^2)^2 + (x^3)^2 - (x^4)^2 - (x^5)^2 = {3\over
\mid\Lambda\mid}. \eqno(3.10)$$
The induced metric satisfies
$$ R_{\alpha\beta} = \Lambda g_{\alpha\beta},\eqno(3.11) $$
and plays the r\^ole of de Sitter spacetime for $SO(2,2;\Bbb R)$. Note
that as a consequence of chronal-chiral symmetry there is only one de
Sitter spacetime for $ SO(2,2;\Bbb R).$ The anti-de Sitter version with
the opposite sign of $\Lambda$ corresponds to changing
$g_{\alpha\beta}$ to $-g_{\alpha\beta}.$

Suppressing one timelike and one spacelike direction exhibits
${SO(2,2;\Bbb R)/\pm1}$ as the conformal group of $\Bbb R^{1,1}$,
2-dimensional Minkowski spacetime whose conformal compactification is
the 2-torus, ${S^1 \times S^1/\pm1}$. If we regard
$\Bbb R^{1,1}$ as the intersection of a null hyperplane in $\Bbb R^{2,2}$
with the light cone of the origin, we can also consider the further
intersection with spacelike, timelike or null hyperplanes in $\Bbb
R^{2,2}$. These intersect $\Bbb R^{1,1}$ in hyperbolic circles; these are
timelike, spacelike or null curves of constant acceleration. Thus
$SO(2,2;\Bbb R)$ plays the same r\^ole in hyperbolic circle geometry as
the Lorentz group $SO(3,1;\Bbb R)$ in the circle geometry of the
2-dimensional Euclidean plane. In other words the set of hyperbolae
corresponds to two copies of ${(AdS)_3/J}$, where $J$ is the antipodal
map, and  $(AdS)_3$ refers to three dimensional Anti-de Sitter spacetime.

At this point we should emphasize that in all dimensions greater than
two, the conformal group ${\text{Conf}}(s,t;\Bbb R)$ of $\Bbb R^{s,t}$ is
locally isomorphic to $SO_0(s+1,t+1;\Bbb R)$ and the conformal
compactifications of $\Bbb R^{s,t}$ correspond to a set of null rays in
$\Bbb R^{s,t}$. What is special about $\Bbb R^{2,2}$ is that we may
identify $\Bbb R^{3,3}$ with its Kleinian metric as the space of two
forms in $\Bbb R^4$ and hence with the Lie algebra of $SO(4-t,t;\Bbb
R)$. The Kleinian metric is invariant under the adjoint action of
$SO(4-t,t;\Bbb R)$ on $so(4-t,t;\Bbb R)$ and so corresponds  to one of
the two quadratic Casimir operators.

Thought of as $so(4-t,t;\Bbb R)$~,~$\Lambda^2(\Bbb R^4)$, the space of
all two-forms in $\Bbb R^4$, has another
natural metric, the negative of the $Ad_{SO(4-t,t;\Bbb R^4)}$ invariant
Killing metric. If $\omega_{\mu\nu}=-\omega_{\nu\mu}$ is a two form,
then this metric is given by
$$ G(\omega,\omega) =
{1\over 2}\omega_{\mu\nu}\omega_{\lambda\rho}g^{\mu\lambda}g^{\nu\rho}=
g_{ab}\omega^a\omega^b, \eqno(3.12)$$
where $g_{\mu\nu}$ is the flat metric with signature $(4-t,t)$ and
$\omega^a$ with $a=1,2\ldots$ are the components of $\omega_{\mu\nu}$
in a suitable basis.

The (negative) Killing metric has signature $(6,0),(3,3)$ or $(2,4)$
if $t=0,1,2$ respectively. Moreover, since we have two metrics $k_{ab}$
and $g_{ab}$ on $\Lambda^2(\Bbb R^4)$, we can form the Hodge star
operator
$$ (\ast)^a{}_b = k^{ac}g_{cb}, \eqno(3.13)$$
from them.
 Thus
$$ (\ast\omega)_{\mu\nu} =
{1\over
2}\epsilon_{\mu\nu\rho\tau}g^{\rho\lambda}
g^{\tau\sigma}\omega_{\lambda\sigma}. \eqno(3.14)$$
Now it is easily seen that $$ \ast\ast = (-)^t, \eqno(3.15)$$
or equivalently
$$ k^{ab}g_{bc} = (-)^t g^{ab}k_{bc}. \eqno(3.16)$$
Thus, if $t=2$ (or $t=0$) the two metrics can be simultaneously
diagonalized to yield the orthogonal direct sum decomposition
$$ SO(2,2;\Bbb R) = SL(2;\Bbb R)_L \oplus SL(2;\Bbb R)_R, \eqno(3.17)$$
where the left and right subspaces $SL(2,\Bbb R)_L$ and $SL(2,\Bbb R)_R$
are  the eigenspaces of the Hodge $\ast$ operator with
eigenvalues $1$ and $-1$ respectively. The eigenspaces are generally
referred to as the self-dual and anti-selfdual spaces.
Restricting each factor to the (negative) Killing metric, they each have
signature $(2,1)$ while the Klein metric has the same or
opposite signature respectively. It follows that for self-dual, or
anti-self-dual two forms, the property of being simple coincides with
that of being null. Such self-dual (or anti-self-dual) simple two forms
are called $\alpha$-(or $\beta$-)forms.
Moreover the simultaneous requirements that $\omega_{\alpha\beta}$ be
simple
$$ \omega_{\alpha\beta}= u_{[\alpha}v_{\beta]}, \eqno(3.18)$$
and self-dual  (or anti-self-dual)
$$ \omega_{\alpha\beta}= \pm {1\over
2}\epsilon_{\alpha\beta}{}^{\gamma\delta}\omega_{\gamma\delta},\eqno(3.19)
$$ which is to say that
$$ u_{[\alpha}v_{\beta]} = \pm {1\over 2}
\epsilon_{\alpha\beta}{}^{\gamma\delta} u_{\gamma}v_{\delta},
\eqno(3.20)$$
are easily seen to imply that the two one-forms $u_{\alpha}$ and
$v_{\beta}$ are themselves both null and orthogonal to each other
$$ g^{\alpha\beta}u_{\alpha}u_{\beta} =
g^{\alpha\beta}v_{\alpha}v_{\beta} = g^{\alpha\beta} u_{\alpha}
v_{\beta} = 0. \eqno(3.21)$$

It follows from the above that the 2-planes annihilated by
$\omega_{\alpha\beta}$ are themselves totally null, that is spanned by
two vectors $a^{\alpha}= g^{\alpha\beta}u_{\alpha}$ and
$b^{\alpha}= g^{\alpha\beta} v_{\beta}$ such that
$$ a^{\alpha}a^{\beta}g_{\alpha\beta} =
b^{\alpha}b^{\beta}g_{\alpha\beta}
= a^{\alpha}b^{\beta}g_{\alpha\beta} =  0, \eqno(3.22) $$
with
$$ \omega_{\alpha\beta} a^{\alpha} = \omega_{\alpha\beta} b^{\alpha} =
0. \eqno(3.23)$$

Since $\alpha$ and $\beta$ planes play an important r\^ ole in what
follows, we shall pause to describe them in more detail.
To do so we reconsider the light cone in $\Bbb R^{2,2}$. If the metric
is given by
$$ ds^2 = dx^+dx^- - dy^+dy^-, \eqno(3.24)$$
then a general null 1-form $k(\omega,\alpha,\beta)$ may be parametrized
as $$ k_{\mu} dx^{\mu} = \omega\sin\alpha(-\sin\beta dx^+ + \cos\beta
dy^+) +\omega\cos\alpha(\cos\beta dx^- - \sin\beta dy^-), \eqno(3.25) $$
with $0 \le \alpha \le 2\pi$~;~$0 \le \beta \le 2\pi$,
whence
$$ k^{\mu}k_{\mu} = k^{\mu} {\partial k_{\mu}\over\partial\alpha} =
               k^{\mu} {\partial k_{\mu}\over\partial\beta} = 0.
\eqno(3.26) $$
Moreover
$$ {\partial k^{\mu}\over\partial\alpha}
{\partial k_{\mu}\over\partial\alpha}
=  {\partial k^{\mu}\over\partial\beta}
{\partial k_{\mu}\over\partial\beta}
=0, \eqno(3.27) $$
and
$$ {\partial k^{\mu}\over\partial\alpha}
{\partial k_{\mu}\over\partial\beta}
= \omega^2. \eqno(3.28)$$
Thus the parameters $\alpha$ and $\beta$ label two null directions on
the set of null rays, and the curves $\alpha=$ constant
$0\le\beta\le2\pi$ or $\beta=$ constant $0\le\alpha\le2\pi$ are
circles which wind around each of the two fundamental generators of the
torus. Since
$$ k \we {\partial k\over\partial\alpha} = -\omega^2 (-\sin\beta
dx^+ + \cos\beta dy^+)\we(\cos\beta dx^- - \cos\alpha dy^-),
\eqno(3.29) $$
and
$$ k \we {\partial k\over\partial\beta}  = \omega^2 (\sin\alpha dx^+
+ \cos\alpha dx^-) \we (\sin\alpha dy^+ + \cos\alpha dy^-), \eqno(3.30)$$
these two families of null vectors lie in two totally null 2-planes
which are called the $\alpha$-planes or $\beta$-planes respectively.
Moreover as $\alpha$ varies, we obtain a circle of $\alpha$-planes,
 and as $\beta$ varies, we obtain a circle of $\beta$-planes.

The structure of $\alpha$ and $\beta$ planes has
important consequences for the behavior of massless fields in
Pl\"ucker 4-space. It means, for example, that a massless particle
with 4-momentum $k_1^{\mu}$ can decay into two other massless
particles with 4-momenta $k_2^{\mu}$ and $k_3^{\mu}$ conserving
momentum, so that $$ k_1^{\mu}= k_2^{\mu} + k_3^{\mu}, \eqno(3.31) $$
provided that both $k_2^{\mu}$ and $k_3^{\mu}$ lie either in
the $\alpha$-plane or the $\beta$-plane passing through $k_1^{\mu}$. In
Minkowski spacetime this would only be possible if $k_1^{\mu}$,
$k_2^{\mu}$ and $k_3^{\mu}$ were mutually parallel. More seriously,
particles can decay into products that are more massive than their
progenitors.

 Another interesting feature of massless fields and totally null 2-planes is
that an arbitrary function of two coordinates spanning such a 2-plane is
automatically a solution of the wave equation. Take, for example,
the totally null 2-plane spanned by $x^+$ and $y^+$. Since the wave
equation is $$ {\partial^2\phi\over\partial x^+\partial
x^-}-{\partial^2\phi\over\partial y^+\partial y^-} = 0, \eqno(3.32) $$
then
$$ \phi = f(x^+,y^+) \eqno(3.33)$$
is automatically a solution of (3.32).
Since such solutions are analogous to the left-movers in $\Bbb R^{1,1}$,
we propose calling such solutions left-handed or right-handed
if they are constant on $\alpha$-planes or $\beta$-planes respectively.
Thus, with our conventions, $f(x^+,y^+)$ is a left-handed because it is
constant on the $\alpha$-plane spanned by $x^-$ and $y^-$. Note that
since both $dx^+\we dy^+$ and $dx^-\we dy^-$ are self-dual 2-forms, it
is the case that an arbitrary function $g(x^-,y^-)$  is also
left-handed. Examples of right-handed solutions would be $h(x^+,y^-)$ or
$k(x^-,y^+)$.

How now do we treat a general solution of the wave equation? We can
express $\phi$ as a Fourier integral, so that
$$ \phi = \int_0^\infty \int_0^{2\pi} \int_0^{2\pi}
\omega d\omega d\alpha
d\beta e^{ik_\mu(\omega,\alpha,\beta)x^\mu} \tilde
f(\omega,\alpha,\beta), \eqno(3.34) $$
where $\omega d\omega d\alpha d\beta$ is the invariant measure on the
null-cone. If we perform the $\omega$ integral, we get a
Whittaker-type  formula
$$ \phi = \int_0^{2\pi} d\alpha \int_0^{2\pi}
d\beta~ f(-\sin\alpha\sin\beta
x^++\sin\alpha\cos\beta y^++\cos\alpha\cos\beta
x^--\cos\alpha\sin\beta y^-,
\alpha,\beta), \eqno(3.35)$$
where
$$ f(\lambda,\alpha,\beta) = \int_0^\infty d\omega \omega \tilde
f(\omega,\alpha,\beta) e^{i\omega\lambda}, \eqno(3.36) $$
and $\tilde f(\omega,\alpha,\beta)$, or equivalently
$f(\lambda,\alpha,\beta)$
is an arbitrary function. If we do the $\beta$ integration first,
we find that
$$ \phi = \int_0^{2\pi} d\alpha F(\alpha), \eqno(3.37) $$
where
$$ F(\alpha) = \int_0^{2\pi} d\beta f(-\sin\alpha\sin\beta
x^++\sin\alpha\cos\beta
y^++\cos\alpha\cos\beta x^--\cos\alpha\sin\beta y^-,\alpha,\beta).
\eqno(3.38) $$

$F(\alpha)$ is a superposition of solutions each of which is constant
along the null direction $k^{\mu}(\omega,\alpha,\beta)$. As $\beta$
varies, the null vectors span an $\alpha$-plane specified by the
fixed value of $\alpha$, that is to say we have a left-handed solution
associated with the $\alpha$-plane spanned by
$k^{\mu}$ and ${\partial k^{\mu}\over\partial\beta}$. Performing the
$\alpha$ integration, we see that the general solution may be
expressed as a superposition of left-handed solutions or, by
interchanging the $\alpha$ and $\beta$ integrations, as a
superposition of right-handed solutions. Note that unlike the situation
in $\Bbb R^{1,1}$ where one needs both the left and right movers, for
$\Bbb R^{2,2}$ one only needs either the left-handed solutions or the
right-handed solutions. The reason is presumably because the space
of unoriented null directions in $\Bbb R^{2,2}$ is connected,
whereas in the case of $\Bbb R^{1,1}$ it consists of two disconnected
points. It would appear that the ability to write general solutions of
the wave equation in terms of solutions of a single handedness is
rather similar to the idea of duality found in string theory, where
a single string scattering process must be represented by a number of
distinct Feynman diagrams.

A more group theoretic description of the $\alpha$-planes and
$\beta$-planes is provided by exhibiting the isomorphism $SO(2,2;\Bbb R)
\cong SL(2,\Bbb R)_L\otimes SL(2,\Bbb R)_R/\pm 1$ by regarding flat
Pl\"ucker 4-spacetime $\Bbb R^{2,2}$ as the space of $2\times2$ real
matrices:
$$  {\bold x} = \pmatrix  x^+ & y^+ \\ y^- & x^-\endpmatrix,\eqno(3.39)
$$  with
metric $$ ds^2 = \det {\bold dx} = dx^+dx^- - dy^+dy^-. \eqno(3.40)$$
Left multiplication of ${\bold x}$ by an element $L$ of $SL(2,\Bbb R)$,
and right multiplication by the transpose of another element $L'$ of
$SL(2,\Bbb R)$ preserves the determinant, and hence the Kleinian metric
on $\Bbb R^{2,2}$. The spin space of $SO(2,2;\Bbb R)$ similarly splits
into the direct sum of two real 2-dimensional spin spaces $ S \oplus S'$
of unprimed and primed spinors $\alpha^A$ and $\beta^{A'} $
respectively. Elements of $S$ (or $S'$) are Majorana-Weyl spinors for
$SO(2,2;\Bbb R)$. A spinor dyad for $S$ ($S'$) is given by two unprimed
(primed) spinors $o^A,\iota^A$ ($\tilde o^{A'},\tilde\iota^{A'}$) such
that $$ o^A \iota^B - o^B\iota^A = -\epsilon^{AB}, \eqno(3.41)$$
and
$$ \tilde o^{A'}\tilde\iota^{B'} - \tilde o^{B'}\tilde\iota^{A'} =
-\tilde\epsilon^{A'B'}, \eqno(3.42)$$
where $\epsilon^{AB}$ and $\tilde\epsilon^{A'B'}$ are the symplectic
2-forms on $S$ and $S'$ defining $SL(2,\Bbb R)$. If we parametrize
elements of $SL(2,\Bbb R)$ by four coordinates subject to one constraint
$$ L = \pmatrix a_+ & b_+ \\ b_- & a_-\endpmatrix, \eqno(3.43)$$
with
$$ det L = 1, \eqno(3.44)$$
we see that we may identify $SL(2,\Bbb R)$ with the bi-invariant Killing
metric with three dimensional anti-de Sitter spacetime $(AdS)_3$. To get
$SO(2,1;\Bbb R)$ we must identify $L$ and $-L$ which corresponds to
factoring $(AdS)_3$ by the antipodal map. Thus $SO_o(2,2;\Bbb R)$ with
its Killing metric may be identified with two copies of $(AdS)_3$
quotiented by the simultaneous action of the antipodal map on each
factor. The Lie algebra of $SO(2,2;\Bbb R)$ may be identified with the
tangent space at the origin and splits as a direct sum as was mentioned
earlier. The adjoint action of $ SO(2,2;\Bbb R)$ decomposes into the
adjoint action of the two $SL(2,\Bbb R)$ factors on their Lie algebras.
This action may be described as follows: we may identify  $SL(2,\Bbb R)$
with its (negative) Killing metric with three dimensional Minkowski
spacetime, the adjoint action being equivalent to the usual action of
the Lorentz group. The non-trivial orbits under the Lorentz group action
then decompose into five strata according to how their tangent vectors
are classified: future timelike, past timelike, future null, past null
and spacelike. Each timelike orbit may be identified with 2-dimensional
hyperbolic space $H^2$ and the spacelike orbits with 2-dimensional de
Sitter spacetime $(dS)_2$.

One may choose a basis $(T,I,S)$ in $SL(2,\Bbb R)$ normalized with
respect to the Killing metric such that
$$ -I^2 = T^2 = S^2 =1, \eqno(3.45)$$
and
$$ TIS=1. \eqno(3.46)$$
It follows that $T,I$ and $S$ mutually anticommute. A convenient
representation is
$$ I = \pmatrix 0&1\\-1&0\endpmatrix,\hskip 0.8cm
   T = \pmatrix 1&0\\0&-1\endpmatrix,\hskip0.8cm
   S = \pmatrix 0&1\\1&0\endpmatrix. \eqno(3.47) $$
The relations (3.45) and (3.46) generate $R(2)$, the algebra of all real
$2 \times 2$ real matrices, or what are sometimes called the \lq\lq
pseudoquaternions."  The relations continue to hold when $I,T$ and $S$
act on $\Bbb R^2$ where $I$ provides a complex structure and $T$ and $S$
are a pair of real structures. Thus, for example, if $I^T$ acts on the
right of the matrix ${\bold x}$, we find that
$$ (x^+,x^-,y^+,y^-) \rightarrow (y^+,-y^-,-x^+,x^-), $$
which shows that if we introduce complex coordinates
$$ z^\pm = x^\pm \pm iy^\pm, \eqno(3.48) $$
the effect of I is to multiply the complex coordinates $z^\pm$ by $i.$
In terms of complex coordinates, the flat Pl\"ucker metric becomes
$$ ds^2 = {1\over 2}(dz^+d\bar z^- + d\bar z^+dz^-), \eqno(3.49) $$
which is manifestly pseudo-K\"ahler. The K\"ahler form is antiselfdual
and equal to to
$$\eqalign{ {1\over 2}I_{\alpha\beta}
dx^\alpha\wedge dx^\beta& = {1\over 2}(dy^-\wedge dx^+
+ dx^-\wedge dy^+),\cr
& = {i\over 4}(dz^+\wedge d\bar z^- + d\bar z^-\wedge dz^+).
}  \eqno(3.50)$$
$I^\alpha{}_\beta$ is an isometry of the flat metric, and because
$I^2=-1$
it leaves no vector fixed. However it may rotate a two-planes worth of
vectors into themselves. Such two-planes are called holomorphic with
respect to the complex structure $I$. These holomorphic two-planes may,
or may not, be null.

The situation with regard to the real structures $T$ and $S$ has some
similarities and some differences. Since $S^2=1$ and antisymmetric
$S_{\alpha\beta}=-S_{\alpha\beta}$
we see that
$$ g(SX,SY) = g(X,Y). \eqno(3.51)$$
Thus $S$ is not an isometry of $g$. Moreover, the positive and negative
eigenspaces of  $S$ are totally null. In other words, associated with $S$
are a pair of $\beta$-planes, which correspond in the Lie algebra to the
intersection of the timelike two-plane orthogonal to $S$ with the
light cone (see figure 3.3). In this basis, such $\beta$-planes
correspond to $$ B_\pm = {1\over 2}(I \pm S), \eqno(3.52)$$
and satisfy
$$ B_\pm^2 = 0. \eqno(3.53)$$
$B_\pm$ is thus a nilpotent element of $SL(2,\Bbb R)$.

We may bring out the analogies rather than the differences between real
structures and complex structures by introducing \lq\lq double numbers"
in place of complex numbers. Just as the complex numbers $\Bbb C$ may be
considered as an algebra over $\Bbb R$ generated by unity and another
element $i$ such that $i^2=-1$, we can consider the commutative and
associative algebra of double numbers $\Bbb E$ generated by a unit
element $\Bbb I$, and another generator $e$ such that $e^2=1$.
We introduce double
number-valued coordinates by
$$ \eqalign{ &w^+ = x^+ +e y^+
\cr &         w^- = x^- -e y^-}, \eqno(3.54)$$
together with their conjugates
$$ \eqalign{ &\bar w^+ = x^+ -e y^+\cr
&             \bar w^- = x^- -e y^-},\eqno(3.55)$$
If $S^T$ acts on the right of the matrix ${\bold x}$, it is equivalent to
multiplication by $e$, just as right multiplication by $I^T$ was
equivalent to multiplication by $i$. In these coordinates the metric
takes
the form
$$ ds^2 = {1\over 2}(dw^+d\bar w^- + d\bar w^+dw^-), \eqno(3.56)$$
where now the analogue of the Ka\"hler form is
$$ \eqalign{ S_{\alpha\beta}dx^\alpha dx^\beta &=
{1\over 2}(dx^+\wedge dy^- + dx^-\wedge dy^+)\cr
&=-{e\over 4}(dw^+\wedge d\bar w^- + d\bar w^+\wedge dw^-)}.
\eqno(3.57)$$ Further analogies between complex numbers and double
numbers are revealed by recalling that with respect to the complex
structure $I^\alpha{}_\beta$, the complex valued 2-form
$ T_{\alpha\beta}+iS_{\alpha\beta}$
is closed and holomorphic.
In fact
$$ dz^+\wedge dz^- = -2(T+iS). \eqno(3.58)$$
Analogously for double numbers
$$ dw^+\wedge dw^- = -2(T+eS). \eqno(3.59)$$

The double numbers $\Bbb E$ do not usually receive a great deal of
attention because they are not a division algebra, the modulus
$z'\bar z'=(x^-)^2 - (y^+)^2$ being non-positive. Moreover as an algebra
$\Bbb E$ splits as a direct sum of two copies of
$\Bbb R$, $\Bbb E=\Bbb R \oplus
\Bbb R$, the two copies being generated by the idempotents
$ 2^{-{1/2}} (1 \pm e)$ whose product vanishes.
Nevertheless, the great formal similarity between them and the complex
numbers, including the ability to introduce an analytic function theory
with an obvious analogue of the Cauchy-Riemann equations makes them very
useful as a calculational tool for ultrahyperbolic geometry since one can
interpret most of the usual formulae of complex and K\"ahler geometry
in terms of double numbers. Moreover, it allows a simple passage
from the general analysis of Plebanski {\it et. al}\cite{13} in terms
of complex metrics, with two independent sets of complex coordinates
to the restriction to the ultrahyperbolic case.

An alternative basis for $SL(2,\Bbb R)$ is obtained in terms of
a non-compact generator $T$ and the two nilpotent generators
$A_\pm$ given by
$$ A_\pm = {1\over 2}(S\pm I) \eqno(3.60)$$
which satisfy
$$ \eqalign{ &A_+^2=A_-^2=0,\cr
             &[A_+,A_-]=T,\cr
             &[T,A_\pm]=\pm2A_\pm.\cr}\eqno(3.61)$$
Thus $A_+$ and $A_-$ are null vectors in the Killing metric.
Let us consider $A_-$ which has a two-by-two matrix representation
$$ A_- = \pmatrix 0&0\\ 1&0\endpmatrix, \eqno(3.62)$$
with
$$ \exp(tA_-)=\pmatrix1&0\\ t&0\endpmatrix,\eqno(3.63)$$
acting on the left of the matrix ${\bold x}$
$$ \exp(tA_-):(x^+,x^-,y^+,y^-) \rightarrow (x^+,x^-+t
y^+,y^+,y^-+tx^+), \eqno(3.64)$$
which stabilizes
the self-dual totally null 2-plane (or $\alpha$-plane), $x^+=y^+=0$.
the action of $A_-$ may be reproduced by introducing coordinates taking
values in the algebra $\Bbb D$ of \lq\lq dual numbers,"
that is the algebra
generated by unity and another element $\epsilon$ satisfying
$\epsilon^2=0$. If
$$ z^{\prime\prime} = y^+ + \epsilon x^-, \ \ \ w^{\prime\prime}=
x^++\epsilon y^-,\eqno(3.65)$$ then
$$ A_- : \ \ z^{\prime\prime}\rightarrow\epsilon z^{\prime\prime},\ \ \
w^{\prime\prime}\rightarrow\epsilon w^{\prime\prime}. \eqno(3.66)$$
Since $\det A_- = 0$, it also stabilizes a primed spinor, $o_{A'}$,
say. The converse is also true, every spinor determines a nilpotent
 generator $A_-$ and an associated totally null 2-plane, or $\alpha$
plane corresponding to the simple antiselfdual 2-form whose spinor
expression is $\epsilon_{AB} o_{A'}o_{B'}$.

To illustrate the utility of complex, dual and double numbers for
ultrahyperbolic geometry, we return to $\Bbb R^6$ thought of as the Lie
algebra of $SO(4-t,t;\Bbb R)$, is equivalent to $\Lambda^2(\Bbb R^4)$.
Recall that this space has two distinct metrics on it, the Kleinian
metric $k_{ab}$ given by equation 3.4 and the Killing metric, $g_{ab}$
given by equation 3.12.
Additionally, there is the Hodge star operator, $\ast$, which since
$$ \ast\ast= (-1)^t, \eqno(3.67) $$
may be regarded as providing a complex structure on $SO(3,1;\Bbb R)$, and
a double structure on $SO(4,\Bbb R)$ and $SO(2,2;\Bbb R)$. If
the combinations
$$  a + i b \ \ \ {\text{ or}}\ \ \ \  a + e b $$
are used
where $ a$ and $ b$ are vectors in $so(3;\Bbb R)$ or
$so(2,1;\Bbb R)$, we may exhibit the following isomorphisms:
$$ \eqalign{ &SO(4,1;\Bbb R) \cong SO(3;\Bbb C), \cr
             &SO(4;\Bbb R) \cong SO(3;\Bbb E), \cr
             &SO(2,2;\Bbb R) \cong SO(2,1;\Bbb E). \cr} \eqno(3.68)$$
If instead we had considered the dual vectors in the combination
$$  a+ \epsilon  b,$$
we would have found an isomorphism between the Euclidean group
and rotations
$$ E(3;\Bbb R) \cong SO(3;\Bbb D). \eqno(3.69) $$

If we identify $SO(3,\Bbb R)$ with the real quaternion algebra and
$SO(2,1;\Bbb R)$ with the real pseudo-quaternion algebra,
then we obtain complex quaternions, \lq\lq bi-quaternions" and what
are sometimes called \lq\lq motors" respectively from the isomorphisms
given in equation (3.68). These were introduced originally by Clifford,
Ball, Study and others to discuss rigid body motion in $H^3, S^3$ and
$\Bbb R^3$. The last case is of considerable technological importance.
The relation of all of these concepts to the geometry of
ultrahyperbolic manifolds and of Euclidean three space is quite
remarkable, and is worthy of study even though the physical significance
of Kleinian geometry seems rather tenuous.

The last case is of considerable physical and practical importance, as
well as being mathematically interesting. The Lie algebra of the
Euclidean group $e(3)$ (i.e. the space of \lq\lq motors") is
6-dimensional. It consists of pairs of 3-vectors $(\bold v,\bold\omega)$
where $\bold v$ is an infinitesimal velocity and $\bold\omega$ an
infinitesimal angular velocity with respect to some origin $O$ in $\Bbb
R^3$. Under a change of origin (i.e. under the adjoint action of a
translation $\bold a$, we find
$$ \bold v \rightarrow \bold v + \bold\omega \times \bold a,
\eqno(3.70)$$ $$ \bold\omega \rightarrow \bold\omega, \eqno(3.71)$$
which leaves invariant both the negative Killing metric
$$ \bold\omega \cdot \bold\omega, \eqno(3.72)$$
and the extra quadratic Casimir
$$ 2 \bold\omega \cdot \bold v. \eqno(3.73)$$
The latter provides the space of motors with a Kleinian metric of
signature $(3,3)$ which coincides with $g^{ab}$ if we regard the
components $(\bold v,\bold\omega)$ of a  motor as giving the six
components $v_a$ of a co-vector in $\Bbb R^{3,3}$. The motors $(\bold
v,\bold\omega)$ and $(\lambda\bold v,\lambda\bold\omega)$ define the same
one-parameter subgroups of $E(3)$ and have as orbits on $\Bbb R^3$
helical curves with an axis (straight line) and pitch
$p=\bold\omega\cdot\bold v/ \bold\omega^2$. For this reason  elements
of the projective space $P(E(3))=P_6(\Bbb R)$ are referred to as \lq\lq
screws".  Note that every Euclidean motor is a twist about a
screw.

The rate of doing work $dW\over dt$ by a system of forces equivalent to
a net force $\bold F$ and a couple $\bold G$ is
$$ {dW\over dt} = \bold v \cdot \bold F + \bold\omega\cdot\bold G.
\eqno(3.74)$$
Moreover under a change of origin we have
$$ \bold F \rightarrow \bold F, \eqno(3.75)$$ and
$$ \bold G \rightarrow \bold G + \bold F \times \bold a, \eqno(3.76)$$
which leaves invariant both the right hand side of equation (3.74) as
well as both
$$ \bold F \cdot \bold F, \eqno(3.77)$$
and
$$ \bold F \cdot \bold G. \eqno(3.78)$$

The rate of doing work $dW\over dt$, that is the right hand side of
(3.74) is invariant under a change of origin, and so we can identify the
pair $(\bold F,\bold G)$, called a \lq\lq wrench" as an element of the
dual space $e(3)^*$. The transformations 3.75 and 3.76 are just the
coadjoint actions of the Euclidean group $E(3)$ on the dual of its Lie
algebra $e(3)^*$. In an appropriate basis the six components of the
wrench $(\bold F,\bold G)$ may be written as $g^a$ so that (3.74) becomes
$$ {dW\over dt} = v_a g^a \eqno(3.79)$$
Note that the components $(\bold y - \bold x , \bold x \times \bold y)$
of the Pl\"ucker coordinates of the line through $\bold x$ and $\bold y$
transform under a change of origin as
$$ \bold y - \bold x \rightarrow \bold y - \bold x \eqno(3.80)$$
$$ \bold x \times \bold y  \rightarrow \bold x \times \bold y + (\bold y
- \bold x)\times \bold a \eqno(3.81)$$
which is the same way as 3.77 and 3.78 transform, that is under the
coadjoint action. A comparison with 3.3 and 3.5 reveals why the
components of a motor should be thought of as comprising a covariant
vector rather than contravariant vector. We could also have adopted a
two-form notation with
$$ G^{ij} = \epsilon^{ijk} G^k, \eqno(3.82)$$
$$ G^{0i} = F^i, \eqno(3.83)$$
$$V_{ij}= \epsilon_{ijk} \omega_k, \eqno(3.84)$$
$$ V_{0i} = v_i. \eqno(3.85)$$
Now equation (3.79) can be rewritten as
$$ {dW\over dt} = {1\over 2} V_{\alpha\beta} G^{\alpha\beta}.
\eqno(3.86)$$

There is obviously a reciprocity or duality between wrenches and motors
in that the pairs $\bold\omega$ and $\bold F$ and $\bold G$ play
identical r\^oles in all formulae. This reciprocity is just the
isomorphism between $\Bbb R^{3,3}$ and its dual space induced by the
Kleinian metric $k_{ab}$. In physicists language this corresponds to the
lowering and raising of indices. Thus we may set
$$ G_{\alpha\beta} = {1\over 2} \epsilon_{\alpha\beta\mu\nu} G^{\mu\nu},
\eqno(3.87)$$
which then has components
$$ G_{0i} = G_i, \eqno(3.88)$$
$$ G_{ij} = \epsilon_{ijk}F_k. \eqno(3.89)$$
Correspondingly $V^{\alpha\beta}$ has components
$$ V^{0i} = \omega^i, \eqno(3.90)$$
$$ V^{ij} = \epsilon^{ijk} V^k. \eqno(3.91)$$

The isomorphism between $e(3)$ and $e(3)^*$ is an essentially a three
dimensional phenomenon. It is not true for $e(n)$ and $e(n)^*$, $n \ne
3$. It arises solely because of the extra Casimir, equation (3.73). It
means, among other things, that all statements in statics (i.e. about
wrenches) have analog statements in the theory of infinitesimal
kinematics (i.e. about motors and screws). Moreover there is also a
connection to line geometry, which in turn has important application in
optics and symplectic geometry, all resulting from  the
ultrahyperbolic metric structure.

It is illuminating to place the geometry of Kleinian 6-space in a
 more general context by looking at it from a slightly different
point of view. Consider a general Lie group $G$ with Lie algebra $\goth
g$. Let $\goth g^*$ be the dual of the Lie algebra $\goth g$ so that if
$\omega^i \in \goth g $ and $ v_i \in \goth g^*$, $i=1,2 \dots
{\text {dim}}G$, there is a natural product which we write as
$$ \langle v,\omega \rangle = v_i \omega^i. \eqno(3.92)$$

The cotangent bundle $T^*(G)$ may, since any Lie group may be
parallelized by, for example, left translating a basis for $\goth g^* =
T_l^*(G)$ over $G$ be identified topologically with $G \times \goth
g^*$. We may also endow $T^*(G)$ with a group structure by considering
the semi-direct product of $G \ltimes \goth g^*$ where $\goth g^*$ is
thought of as an additive abelian subgroup, and $G$ acts on $\goth g^*$
by the co-adjoint action. The Lie algebra of $G \ltimes \goth g^*$ may be
identified with the direct sum $ \goth g \oplus \goth g^*$. The
cotangent bundle $T^*(G)$ is naturally a symplectic manifold with a
symplectic form, at the origin, given by
$$ \Omega = dv_i \wedge d\omega^i. \eqno(3.93)$$
One can also endow $\goth g \oplus \goth g^*$ with a Kleinian metric,
$g$, using the product given by equation (3.92)
$$ g = 2 v_i \omega^i. \eqno(3.94)$$
Using left translation we can extend this metric over all of $T^*(G)$.
Moreover, as the metric and symplectic structures are compatible we can
view $T^*(G)$ as a pseudo-K\"ahler manifold with complex structure at the
origin given by
$$ J(\omega) = \omega, \eqno(3.95)$$
and
$$ J(V) = -V. \eqno(3.96)$$
Thus in the obvious basis for $\goth g \oplus \goth g^*$
$$ g = \pmatrix 0&1\\1&0\endpmatrix, \eqno(3.97)$$
$$ J = \pmatrix 1&0\\0&-1\endpmatrix, \eqno(3.98)$$
$$ \omega = \pmatrix 0&1\\1&0\endpmatrix. \eqno(3.99)$$
The theory given above is quite general and may well be useful in, for
example, quantizing a particle moving on a group manifold. The particular
case of interest is when $G=SO(3)$ in which case $T^*(G) =
 SO(3) \ltimes \Bbb R^3 = E(3) $, the Euclidean group, and we are back
to the case of the theory of \lq\lq motors," $\omega$ being an
infinitesimal angular velocity and $v$ and infinitesimal velocity.

\vskip1cm
{\bf \S 4 Curved Kleinian Einstein Metrics}
\vskip0.5cm
In this section we wish to make a few remarks about curved Kleinian
4-metrics. We commence by noticing that on topological grounds not every
4-manifold $\Cal M$ can admit such a metric. Every paracompact manifold
$\Cal M$ admits a (not unique) Riemannian metric, $g_{ab}$ say,  and thus
if $\Cal M$ admits a Kleinian metric $k_{ab}$, then we may diagonalize
$k_{ab}$ relative to $g_{ab}$. it follows that $\Cal M$ must admit an
everywhere non-vanishing 2-plane distribution. According to Atiyah and
Dupont \cite{14} a necessary condition for this, assuming that $\Cal M$
is closed, $\partial\Cal M = \emptyset$, and orientable, and the 2-plane
distribution to be oriented, is that the Euler character $\chi$ and
the signature $\tau$ obey certain conditions, namely
$$ \chi = 0~~mod~~2, \eqno(4.1)$$
and
$$ \tau = \chi~~mod~~4. \eqno(4.2)$$
Hence only if (4.1) and (4.2) are satisfied can a manifold admit a
Kleinian metric.

Let us now turn to metrics
that obey the Einstein   condition
$$ R_{ab}=\Lambda g_{ab}. \eqno(4.3) $$
Whilst this is not directly relevant to string theory, it is useful to
examine such metrics as they illustrate some of the peculiarities of
Kleinian geometry. The first obvious non-flat example is the constant
curvature space, which is a generalization of de Sitter spacetime.
Consider the quadric in $\Bbb R^{2,3}$ given by
$$ (X^1)^2 + (X^2)^2 + (X^3)^2 - (X^4)^2 - (X^5)^2 = {3\over \Lambda},
\eqno(4.4)$$
with $\Lambda > 0$. The isometry group on the surface is
$SO(3,2;\Bbb R)$.
This is same
construction as for Lorentzian anti-de Sitter space except that the
cosmological constant has been chosen to be positive rather
than negative. We should however note that in contrast to the Lorentzian
case there is just a single metric of
constant curvature. This is because changing the sign of $\Lambda$ is
the same as changing the overall sign of the metric, which is a discrete
isometry in the present case. One could regard (4.4) as a generalized
Friedmann-Robertson-Walker metric in a variety of ways, rather
as one does for de Sitter spacetime, but we will not explore this here.

We now pass on to the analog of the Schwarzschild vacuum
solution. For Lorentzian
signature,  the Schwarzschild metric is usually regarded as the
field of a particle with a timelike worldline. Thus it has as its
isometry group $SO(3) \otimes \Bbb R$, the $SO(3)$ having
two-dimensional orbits, and the isometry group being the stabilizer of a
timelike line in Minkowski spacetime. This property completely
characterizes the metric up to the sign of the mass parameter $M$. Both
possible signs give rise to an incomplete singular metric. If $M>0$, the
singularity is hidden inside the event horizon, whereas if $M<0$ there
would be a naked singularity. If instead we had wished to consider the
gravitational field of a tachyon, that is a particle with a
spacelike worldline, we would replace $SO(3)$ by $SO(2,1)$.
We then have a choice as to whether the orbit of $SO(2,1)$ is spacelike,
giving the two-dimensional hyperboloid $H^2$,
or timelike, giving
two-dimensional de Sitter spacetime $(dS)_2$. We also have a choice
of whether the mass is positive or negative. Since these metrics are
warped products of a two-dimensional Lorentzian spacetime and a
Riemannian two-surface, we may illustrate the possibilities by he
self-explanatory diagrams below which include the analogous possibilities
for Kleinian and Riemannian choices of signature.

It is noteworthy that among the list are three complete
non-singular metrics, one
for each signature. In each of these cases, the $(r,t)$ plane
has a positive definite
metric, $t$ being periodic with period $8\pi M$, and $M>0$.

We can construct these solutions in a way that is analogous to the de
Sitter space example. They can be regarded as algebraic varieties in a
seven dimensional flat space of appropriate signature. Lets start with
the Schwarzschild solution of Lorentz signature. It can be regarded as
an algebraic variety in $\Bbb R^{6,1}$. Thus consider the space
$$ ds^2 = (dX^1)^2 + (dX^2)^2 + (dX^3)^2 + (dX^4)^2 + (dX^5)^2
+ (dX^6)^2 - (dX^7)^2 \eqno(4.5) $$
and construct the four-dimensional submanifold formed by the
intersection of the following three hypersurfaces \cite{15}
$$ \eqalign{ &(X^6)^2 -
(X^7)^2 + {4/3}(X^5)^2 = 16M^2, \cr &((X^1)^2 + (X^2)^2 + (X^3)^2)(X^5)^4
= 576M^6, \cr & \sqrt{3} X^4X^5 + (X^5)^2 = 24M^2. }\eqno(4.6)$$
We can relate this to the usual exterior Schwarzschild solution by
means of the following substitutions:
$$ \eqalign{ &X^1 = r\sin\theta\cos\phi, \cr
          &X^2 = r\sin\theta\sin\phi, \cr
          &X^3 = r\cos\theta, \cr
          &X^4 = -2M(2M/r)^{1\over 2} + 4M(r/2M)^{1\over 2}, \cr
          &X^5 = (24M^3/r)^{1\over 2}, \cr
          &X^6 = 4M(1-2M/r)^{1\over 2} \cosh(t/4M),\cr
          &X^7 = 4M(1-2M/r)^{1\over 2} \sinh(t/4M). }\eqno(4.7) $$
In fact, the algebraic variety described here does in fact cover the
entire Kruskal manifold. Since this construction works in a flat seven
dimensional space, it is clear that we can construct metrics of other
signatures in a straightforward way by complexifying the entire
construction, and taking real sections in an appropriate way. By this
technique, we can find all of the metrics alluded to earlier. We will
not do this explicitly here as the method is fairly transparent, but
rather tedious to record.

\vskip1cm
{\bf \S 5 Self-dual metrics}
\vskip0.5cm
The requirement that the curvature of a 4-metric $g_{ab}$ of signature
$(2,2)$ be self-dual,
$$ R_{ab}{}^{cd}={1\over 2}\epsilon_{ab}{}^{ef}R_{cdef}, \eqno(5.1)$$
is easily seen to imply that the metric is Ricci flat and has holonomy
$SL(2,\Bbb R)$ \cite{16}. Such metrics are also said to be half-flat. In
$4n$- dimensions, metrics with Kleinian signature and holonomy $Sp(2n,\Bbb
R)$ have been termed hypersymplectic by Hitchin \cite{17}. Hence self-dual
Kleinian metrics are hypersymplectic. In
two component notation, the only non-vanishing components of the Weyl
curvature spinor is $\Psi_{ABCD}$. Thus there exist a pair of
covariantly constant spinor fields $\iota^{A'}$ and $o^{A'}$, which may
be normalized conveniently so that
$$ o^{A'}\iota_{A'} = 1 \Leftrightarrow \delta^{A'}_{B'} = -\iota^{A'}
o_{B'} + o^{A'}\iota_{B'}. \eqno(5.2) $$
The bundle of anti-self-dual 2-forms is flat  and one may choose a
basis such that
$$ A_+ = \iota_{A'}\iota_{B'} = {1\over 2}(S+I), \eqno(5.3)$$
$$ A_- = -o_{A'}o_{B'} = {1\over 2}(S-I), \eqno(5.4)$$
$$ T = -o_{A'}\iota_{B'} - \iota_{A'}o_{B'}. \eqno(5.5)$$
Hence
$$ I = o_{A'}o_{B'} - \iota_{A'}\iota_{B'}, \eqno(5.6)$$
$$ S =-o_{A'}o_{B'} - \iota_{A'}\iota_{B'}. \eqno(5.7)$$
Thus $I,S$ and $T$
are covariantly constant. It follows that $I$ endows $\Cal M$ with the
structure of a complex K\"ahler manifold, as indeed does every other
2-form lying in the 2-sheeted hyperbola in $\Lambda_-^2(\Cal M)$.
The same statements can be made but with \lq\lq complex" changed to
\lq\lq double" and \lq\lq double-sheeted" to \lq\lq single-sheeted."
Moreover, $A_+$ and $A_-$ belong to two circles worth of \lq\lq dual"
solutions, i.e. to a single circles worth of covariantly constant
$\beta$-planes.

Let us consider now a self-dual Kleinian four manifold admitting an
isometry group $G$ generated by Killing vector fields $K^A$. If $I,S$
and $T$ are invariant under the action $G$ then $G$ is said  to act
(locally) hypersymplectically. The action of $G$ is holomorphic with
respect to every complex structure on the double sheeted hyperboloid,
and similar statements apply in terms of double and dual structures.
The necessary and sufficient conditions that the action is (locally)
hypersymplectic
 is that
$$ {\Cal L}_K{}\{I,S,T\} = 0 \eqno(5.8)$$
for every Killing vector filed $K \in \goth g$, the Lie algebra of $G$.
since $I,S$ and $T$     are covariantly constant we obtain as a
necessary and sufficient condition for hypersymplecticity is the the
two-form $\nabla_aK_b$ be self-dual. One then refers to the Killing
vector being self-dual. If the action of $G$ does not leave $I,S$ and
$T$ invariant, it must act on them preserving the Killing metric, in
other words       the action of $G$ provides a homomorphism from $G$  to
a (possibly improper) subgroup of $SO(2,1;\Bbb R)$. For example, if $G$
is one-dimensional, its action on the self-dual two form may either be
a spacelike rotation leaving invariant a privileged complex structure,
a Lorentz boost leaving invariant a privileged double structure, or a
null rotation leaving invariant a privileged dual structure. In the
spacelike case it is clear that $G$ will not leave invariant any dual
or double structure, while in the second case it will leave invariant
two privileged dual structures - or equivalently two privileged
foliations by $\beta$-planes, while in the last case there will be no
privileged double or complex structures.

The classification of isometries  just given is very similar to that in
the positive definite case. Moreover it is obvious that by means of
analytic continuation (or \lq\lq Wick" rotation) of known examples of
Hyper-K\"ahler manifolds one may obtain numerous local forms of
hypersymplectic   4-manifolds. Since all known Hyper-K\"ahler
4-manifolds admit isometries, the corresponding hypersymplectic forms
will also admit isometries. The analytically continued Killing fields
$K^a$ will always have, however, non-vanishing length $g_{ab}K^aK^b$.
 It follows that  various constructions  and uniqueness arguments in
the Riemannian case will pass over to the Kleinian case without much
change. However, there is a special case which cannot occur in
Riemannian geometry which is when the Killing vector field is null,
$$ g_{ab}K^aK^b = 0. \eqno(5.9)$$
This specific case will be studied in the next section. Now let us turn
our attention to the case of a self-dual Killing vector in a self-dual
spacetime where $K$ is not null. The results of Tod and Ward show that
it may be cast in the form
$$ ds^2 =  V^{-1}(d\tau + \omega_idx^i)^2 - V\eta_{ij}dx^idx^j
 \eqno(5.10)$$
where
$$ curl~\omega = grad~ V   \eqno(5.11)$$
with $grad$ and $curl$ being defined with respect to the flat
Lorentzian three metric $\eta_{ij}$. This of course implies that flat
space Laplacian acting on $V$ must vanish.  Similar remarks apply in the
case of a non-self-dual Killing vector in a self-dual space as
has been considered by Boyer and Finley \cite{18}, and Gegenberg and Das
\cite{19}, and also Park \cite{20}.

If instead we had considered the Riemannian form of these metrics,
$$ ds^2 = V^{-1}(d\tau + \omega_idx^i)^2 + V\delta_{ij}dx^idx^j,
\eqno(5.12)$$ we would have found a family of Hyper-k\"ahler metrics
that have a self-dual Killing vector and self-dual curvature,
and that are Asymptotically Locally Euclidean (ALE).
These are the "N-center" metrics where
$$ V = \sum_i^N {1\over \mid \bold x - \bold x_i \mid}, \eqno(5.13)$$
with $\bold x_i$ is a Euclidean three vector, and $\omega$ is
determined by equation (5.11). If $N=1$, then the metric is that of flat
space. If $N=2$, the space is the Eguchi-Hanson instanton metric.
In general the N-center metrics are the unique self-dual metrics
that ALE and whose boundary at infinity is $S^3\over Z_2$.
One can analytically continue any of these metrics to find
hypersymplectic Kleinian metrics, however there is a considerable
arbitrariness in how one goes about this. We now give an example which
appears to have some stringy significance.

Consider the Eguchi-Hanson metric \cite{21} in the form
$$ ds^2 = {dr^2\over f^2} + {1/4}r^2(\sigma_1^2 + \sigma_2^2 +
f^2\sigma_3^2), \eqno(5.14)$$
where
$$ f^2 = (1-{a^4\over r^4}), \eqno(5.15)$$
and $\sigma_i$ are a set of left-invariant one forms  on $S^3$.
Thus
$$ d\sigma_i = -{1\over 2}\epsilon_{ijk}\sigma_j
\wedge\sigma_k. \eqno(5.16)$$
Explicitly, in terms of Euler angles $(\theta,\phi,\psi)$,
$$ \sigma_1 + i \sigma_2 = e^{-i\psi}(d\theta + i \sin\theta d\phi),
\eqno(5.17)$$
$$ \sigma_3 = d\psi+\cos\theta d\phi. \eqno(5.18)$$
One way in which one can analytically continue this metric is by the
coordinate transformation $\theta \rightarrow i\theta$,
so that the metric becomes
$$ ds^2 = {dr^2\over f^2} +{1\over 4}r^2(-d\theta^2 - \sinh^2\theta
d\phi^2 + f^2(d\psi + \cosh\theta d\phi)^2). \eqno(5.19)$$
This can be rewritten in terms of a pseudo-orthonormal basis of
one-forms
$$ e^0={dr\over f},\ \ \ \  e^1={1\over 2}r\sigma_1, \ \ \ \  e^2={1\over
2}r\sigma_2, \ \ \ \ e^3= {1\over 2}rf\sigma_3, \eqno(5.20)$$
where now $\sigma_i$ are a basis of $SO(2,1;\Bbb R)$ invariant
one-forms:
$$\sigma_1 = \cos\psi d\theta + \sin\psi\sinh\theta d\phi, \eqno(5.21)$$
$$\sigma_2 = -\sin\psi d\theta + \cos\psi\sinh\theta d\phi, \eqno(5.22)$$
$$\sigma_3 = d\psi + \cosh\theta d\phi. \eqno(5.23)$$
Hence
$$ d\sigma_i = -{1\over 2}c_{ijk}\sigma_j\wedge\sigma_k.
 \eqno(5.24)$$
The $c_{ijk}$ being the structure constants of $so(2,1)$
The above metric is hypersymplectic as can be seen by constructing the
basis of self-dual 2-forms described earlier. Explicitly
$$ I = e^0\wedge e^3 - e^1\wedge e^2, \eqno(5.25)$$
$$ S = e^0\wedge e^2 + e^1\wedge e^3, \eqno(5.26)$$
$$ T = e^0\wedge e^1 + e^2\wedge e^3. \eqno(5.27)$$
This  metric has the following stringy interpretation. Consider the
submanifold spanned by $e^1$ and $e^2$. This two surface is Riemannian
and of constant curvature. It therefore is conformal to any Riemann
surface of genus $g>1$, after appropriate identification to find the
unit cell.
Let us suppose that this is the string worldsheet
$\Sigma$, appropriate for string loop diagrams. It has Riemannian
signature indicating that it corresponds to a classically forbidden
process, but nevertheless one that is quantum mechanically allowed.
The four-manifold is then a self-dual metric
on $T^\ast(\Sigma)$. This is precisely the situation envisaged by
Ooguri and Vafa. This lends some support to the conjecture that $N=2$
string theory is in fact the same thing as self-dual gravity.

A second rather different continuation can be found by making the
analytic continuation $\phi \rightarrow i\phi,~~ \psi \rightarrow
i\psi$.
Were it the case that $f=1$, then this continuation would yield a
space diffeomorphic to the previous manifold. However, we instead find
that
$$ ds^2 = {dr^2\over f^2} +{1\over 4}r^2\bigl(d\theta^2 - \sin^2\theta
d\phi^2 - f^2(d\psi+\cos\theta d\phi)^2\bigr). \eqno(5.28)$$
This metric is also hypersymplectic. Now however the analog of the
subspace spanned by $e^1$ and $e^2$ is a Lorentz manifold
rather than a  Riemannian one, again of constant negative curvature,
therefore it represents a physical string worldsheet $\Sigma$.,
corresponding to a classically allowed process. Despite this difference,
the four manifold  still has a Kleinian self-dual metric and is  of the
form $T^\ast(\Sigma)$. It would therefore seem that Kleinian self-dual
four metrics provide an arena for discussing both quantum and classical
processes on a democratic footing. This lends further support to the
idea that string theory is in some, as yet ill-understood way, related to
self-dual gravity.

\vskip 1cm
{\bf \S 6 Cosmic Strings}
\vskip0.5cm

There has been some considerable interest recently in the
relation between on the one hand cosmic strings and solitons, and on the
other hand fundamental strings and {\it p}-branes. In particular
\cite{22},  a fundamental string was modelled by a solution of the
equations of motion of the zero-slope limit of the superstring, with a
distributional source with support on the worldsheet of the string. On
the other hand, there is a dual formulation in which the ten-dimensional
superstring written in terms  of seven-form field strengths, there are
source-free soliton like solutions of the zero-slope equations of motion
representing a 5-brane. In the present section, we shall show how to
construct non-singular, source-free solutions of the self-dual Einstein
equations with signature $(2,2)$ which may be interpreted as a \lq\lq
thickening" or \lq\lq de-singularization" of a distributional cosmic
string with curvature supported on a totally null 2-plane. We shall
begin by describing the distributional model, and then its thickening.  The
idea of the distributional model is due to Mason \cite{23}, but our specific
construction is a little different.

A distributional \lq\lq Regge Calculus" model of a cosmic string may
be obtained by taking a flat spacetime ${\Bbb R^{t,s}}$, with $s+t=4$, and an
isometry, following the method of Ellis and Schmidt \cite{24}. A suitable
isometry $\Gamma$ is one with an 2-dimensional fixed point set $\Cal M^2$ and
which lies in a one-parameter subgroup. For the construction, it is necessary
to fix a particular parameter $c$, so that

$$ \Gamma : x^\alpha \rightarrow (\exp cM^\alpha_\beta) x^\beta,
\eqno(6.1)$$ For example, in Euclidean space this fixes the rotation angle,
which is otherwise ambiguous by the addition of multiples of $2\pi$.

The element $M^\alpha_\beta$ of the Lie algebra is determined up to scale by
the fixed point set, and in fact $\Cal M^2$ is given by points $x^\alpha$ such
that $$ M^\alpha_\beta x^\beta = 0.  \eqno(6.2)$$

The spacetime is obtained by an identification of points on the universal
covering space of ${\Bbb R^{t,s}}$ with the
fixed point set $\Cal M^2$ removed. This covering space is denoted
$\Cal C$.

The action of $\Gamma$ lifts uniquely to the covering space, due to the choice
of parameter in $(6.1)$. The fundamental group of $({\Bbb R^{t,s}} - \Cal M^2)$
also acts freely on the covering space. Let $e$ be a generator of the
fundamental group, and identify points under the action of
$\Gamma$ composed with $e$, so that the exterior of the cosmic string is given
by $\Cal M^4 = \Cal C/(\Gamma\circ e)$.

The resulting spacetime is homeomorphic to ${\Bbb R^{t,s}}-\Cal M^2$, but has a
non-standard metric, the holonomy group for curves encircling the string
worldsheet $\Cal M^2$ being generated by $\Gamma$. The metric varies
continuously with the parameter $c$, the case $c=0$ being flat space. The
points $\Cal M^2$ can be reinserted if desired.

The example we are interested in is the case of $\Bbb
R^{2,2}$, with $\Gamma$ an isometry fixing a totally null 2-plane.
By contrast, the example due to Mason, which is constructed by identifying
$\Bbb R^{2,2}$ by $\Gamma$ alone,
results in a rather bizarre spacetime. If the points $\Cal M^2$
are left in, the identification is non-Hausdorff (but T1). If the points of
$\Cal M^2$ are removed, the spacetime is Hausdorff but with homotopy type
$S^1\times S^1$. This is somewhat reminiscent of Misner's spacetime
\cite{25} with $t=1$, though the later remains non-Hausdorff even when $\Cal
M^2$ is removed.

Now return to our construction, which includes the action of $\pi_1$. Suppose
the string worldsheet $\Cal M^2$ lies on $x_1=x_2=0$, and that
$M_{\alpha\beta}$ is scaled so that $M_{12}=1$.
The distributional Riemann curvature tensor has support on the string
worldsheet $\Cal M^2$ and is given by

$$ R_{\alpha\beta\gamma\delta} =
cM_{\alpha\beta}M_{\gamma\delta} \delta(x_1)\delta(x_2), \eqno(6.3)$$
as follows from the definition of the Riemann tensor in \cite{26}.

The distributional Ricci tensor associated with the cosmic string is
given by
$$ R_{\alpha\beta} = c M_\alpha^\sigma M_{\sigma\beta} \delta(x_1)\delta(x_2).
\eqno(6.4). $$

Clearly the string is therefore entirely specified by the simple 2-form
$M_{\alpha\beta}$ and the strength of the source $c$.

For conventional cosmic strings, $t=1$ and $s=3$, and $\Gamma$ is a
spacelike rotation through $c$ about an axis lying in the
string, and $c$ is the deficit angle. Thus
$$ \exp(cM) = \pmatrix 1 &0 &0 &0 \\ 0 &\cos c&
-\sin c &0 \\ 0 &\sin c &\cos c &0 \\ 0
&0 &0 &1 \endpmatrix. \eqno(6.5)$$
The string worldsheet is the timelike 2-surface
$x_1=x_2=0$. The Ricci tensor is non-zero.
The case of null strings in Minkowski spacetime has been discussed by Bruno,
Shapley and Ellis \cite{27}. They take
$$ M^\alpha{}_\beta = \pmatrix 0 &0 &0 &1\\0 &0 &0 &1\\0& 0 &0 &0 \\1 &-1
&0 &0 \endpmatrix. \eqno(6.6) $$
The null string worldsheet satisfies
$$ x^0-x^1 = 0, \eqno(6.7) $$
$$ x^3 = 0. \eqno(6.8)$$
Since
$$ M^\alpha{}_\beta M^\beta{}_\gamma = \pmatrix 1& -1& 0 &0\\ 1& -1& 0
&0\\ 0 &0 &0 &0 \\0 &0 &0 &0\endpmatrix, \eqno(6.9)$$
the distributional Ricci curvature is not zero.
Thus in Minkowski spacetime neither timelike or null strings are
distributional source-free solutions of the Einstein vacuum equations,
and one does not expect to find non-singular thickenings which are
Ricci-flat either.

By contrast in $\Bbb R^{2,2}$, Mason  pointed out that
a simple self-dual (or anti-self-dual) 2-form $M_\alpha{}_\beta$ will
satisfy
$$ M^\alpha{}_\beta M^\beta{}_\gamma = 0. \eqno(6.10)$$
and therefore the associated cosmic string has distributional
curvature with support on an $\alpha$-plane (or $\beta$-plane if
anti-self-dual). An explicit example is provided   by introducing
coordinates such that
$$ ds^2 = dx^+ dx^- - dy^+ dy^- \eqno(6.11)$$
The null self-dual $O(2,2)$ transformations, $\Gamma$, are
$$ \eqalign{& x^+ \rightarrow x^+, \cr
            & y^+ \rightarrow y^+, \cr
            & x^- \rightarrow x^- + c x^+,\cr
            & y^- \rightarrow y^- + c y^+,\cr}\eqno(6.12)$$
and these fix the null self-dual two-plane given by
$$ \eqalign{ &x^+ = 0,\cr &y^+=0.\cr} \eqno(6.13)$$
In this (non-orthonormal) basis
$$ M^\alpha{}_\beta = \pmatrix 0 &0 &0 &0\\ 0 &0 &1& 0\\ 0& 0& 0& 0\\1&0& 0
&0\endpmatrix \eqno(6.14)$$
where
$$ M^\alpha{}_\beta M^\beta{}_\gamma = 0.\eqno(6.15)$$
This establishes that the distributional Ricci tensor is indeed zero.
The $O(2,2)$ transformations $\Gamma$ commute with
\roster
\item anti-self-dual rotations which act transitively on
the quadrics $x^+ x^- - y^+ y^- = \text {constant} $
\item two null and covariantly constant translations generated by
$\partial\over\partial x_-$ and $\partial\over\partial y^-$.
\endroster

Since $\Gamma$ is obtained by exponentiating the null and self-dual
Killing vector field $K= x^+{\partial\over\partial y^-} +
y^+{\partial\over\partial x^-}$, the manifold
$$\Cal M^4 = \Cal C/(\Gamma\circ e)  \eqno(6.16)$$
admits a six parameter group isometries which act transitively. It is
perhaps surprising that removing the worldsheet from $\Bbb R^{2,2}$
leaves a spacetime that is still homogeneous.

In searching for a thickening of the cosmic string, it is natural to
insist that the resulting spacetime should continue to admit some of
these symmetries. In particular we shall demand that $\Cal M^4$ admits a
null self-dual Killing vector field.

We therefore have the following:
\proclaim{Proposition}
Every $(2,2)$ self-dual spacetime admitting a null self-dual Killing
vector field $K$ is automatically Ricci flat, and locally be cast into
the following form:
$$ ds^2 = dp dt - {1\over 2}p^2 du( dv+ H(p,u)du) $$
where $H(p,u)$ is an arbitrary $C^2$ function of its arguments and
$K^\alpha \partial_\alpha = \partial_v. $
\endproclaim

Suppose that the one-form associated with the Killing vector is
$\Cal K$, then $d\Cal K$ is self-dual,
$$ \Cal K \wedge d\Cal K = \ast d(g(\Cal K,\Cal K)) \eqno(6.17) $$
and so  since $\Cal K$ is null, it follows that $\Cal K$ is
hypersurface orthogonal. Hence we can put $K^\alpha\partial_\alpha =
\partial_v$ and $\Cal K=K_\alpha dx^\alpha = -2\omega du$, with $v$
being a null coordinate. The  K-invariant metric then takes the form
$$ ds^2 =  P^{-2}(x,y,u)(dx^2 - dy^2) - 2du(\omega dv + m_idx^i + Hdu).
\eqno(6.18)$$
There is a considerable amount of coordinate freedom that preserves this
metric form. Firstly however, we must impose the self-duality of
$d\Cal K$, and hence obtain
$$ \omega = \omega(x-y,u) \eqno(6.19)$$
Suppose that $\omega$ is not only a function of $u$, then setting
$s=\omega $, and $t=x+y$ we obtain
$$ ds^2 = P^{-2}(s,t,u)dsdt - 2sdu(dv + m_s ds +m_tdt + Hdu).
\eqno(6.20)$$
By changing the $v$ coordinate by $v \rightarrow v + g(s,t,u)$ we can
set $m_s = 0$. Now imposing the self-duality of the curvature form
we discover that the metric must be of the form
$$ ds^2 = s^{-1/2}dsdt - 2sdu(dv + H(s,u)), \eqno(6.21) $$
where $H(s,u)$ is an arbitrary function. Now by rescaling $s={1\over 4}
p^2$ we obtain the metric form given. The condition that $H(p,u)$ be
$C^2$ emerges from the requirement that the curvature form be well
defined. A special case not covered by this treatment is when
 $\omega$ is a function of $u$ only. Then $d\Cal K$ vanishes, and the
Killing vector $K$ is covariantly constant. This class of metrics
is included in the form (6.20). \qed

We can now proceed with our cosmic string calculation. The null
self-dual Killing vector field is
$$ K = {\partial\over\partial v}. \eqno(6.22) $$
If one uses the frame basis of 1-forms
$$ e^1 ={1\over 2}dp,\ \ \ \ e^2=dt,\ \ \ \ e^3{1\over
4}p^2du,\ \ \ \ e^4=-(dv+Hdu), \eqno(6.23)$$ in
$$ ds^2 = e^1\otimes e^2 + e^2 \otimes e^1 + e^3\otimes e^4
+ e^4\otimes e^3, \eqno(6.24)$$
then the only non-vanishing component of the curvature 2-form is
$$ \Omega_{13} = {16\over p^2} {\partial\over\partial
p}\bigl(p^2{\partial H\over \partial p}\bigr). \eqno(6.25)$$

Thus the holonomy is not only self-dual, but null. Clearly the metric
will be flat away from $p=0$ if
$$ H = {f^{\prime\prime}\over p}, \eqno(6.26)$$
where prime denotes the derivative with respect to $u$, and $f(u)$
is an arbitrary function of $u$. One can recover the usual flat form of
the metric  by setting
$$\eqalign{
 x^-&= {p\over 2}u(v+{f(u)\over p}) + t - f(u), \cr
 x^+&=p, \cr
 y^-&={p\over 2}(v+({f^\prime(u)\over p}), \cr
 y^+&=pu. } \eqno(6.27)$$
It now follows that the the null Killing vector
${\partial\over\partial v}$
generates the null self-dual rotation $v \rightarrow v+c$
or
$$ \eqalign{ x^- &\rightarrow x^- + cy^+, \cr
           y^- &\rightarrow y^- + cx^+}, \eqno(6.28)$$
and the covariantly constant null Killing vector
${\partial\over\partial t}$ the null translation $t\rightarrow t + e$
or
$$ y^- \rightarrow y^- + e. \eqno(6.29)$$

Thus $\partial\over\partial v$ corresponds to $
y^+{\partial\over\partial x^-}+x^+{\partial\over\partial y^-}$ and
$\partial\over\partial t$ to $\partial\over\partial y^-$. It is
interesting      that the assumption of a null self-dual Killing vector
always implies the existence of an additional null covariantly constant
Killing vector $\partial\over\partial t$. These solutions are thus
members of Plebanski's class of complex solutions \cite{13} to the
vacuum Einstein equations that are analogs of pp-waves. They have also
previously been described by Ward \cite{28} in the context of twistor
theory.

In order to obtain   the cosmic string, one should note that
$$ {d\over dp}(p^2{d\over dp}({1\over\mid p \mid})) = {-1\over
4\pi}\delta(p). \eqno(6.30)$$
Therefore choosing $H={c\over\mid p \mid}$, $c$ constant, we find that
the coordinate change to the flat coordinates $x^+,x^-,y^+,y^-$ becomes
$$\eqalign{ x^-&= {puv\over 2} + {c\over 2} sgn(p) + t -{cu^2\over 2},
\cr x^+&=p, \cr y^-&=pv+{c\over 2}sgn(p), \cr y^+&=pu.}
\eqno(6.31)$$ These coordinate transformations are discontinuous across
the totally null 2-surface $p=0$ i.e.
$$\eqalign{ x^+&=0, \cr y^+&=0. } \eqno(6.32) $$
Points with the same $u,v,$ and $t$ values but with $p$ very small and
positive, and points with $p$ very small and negative are mapped   to
points with identical $x^+$ and $y^+$ but with $x^-$ and $y^-$ which
differ by $c$. The metric (6.21) with $H={c\over\mid p\mid}$ corresponds
to the distributional cosmic string. To obtain a thickened non-singular
string we need only to replace $H$ by a function which is everywhere
bounded, but which tends to $c\over\mid p\mid$ for large values of $p$.
Thus we can construct an everywhere non-singular self-dual spacetime
representing a cosmic string. This contrasts sharply  with the work of
Bruno, Shapley and Ellis \cite{27} who found that in signature $(1,3)$
that they could not eliminate the singularities. In fact the situation is
strongly resembles that for self-dual metrics with signature $(0,4)$.
The \lq\lq orbifold" construction is obtained by quotienting $\Bbb R^4$,
thought of as $\Bbb C^2$ by the action of a discrete subgroup $\Gamma
\subset SU(2)_L \subset SU(2)_L \times SU(2)_R \equiv \Tilde SO(4)$.
The subgroup $\Gamma$ can be thought of as generating a self-dual
rotation acting on $\Bbb R^4$ which leaves fixed the origin. The
resulting orbifold has a Kleinian singularity at the origin. Deletion of
the origin gives a metric with a flat connection but with self-dual
holonomy. Since the holonomy is self-dual any distributional curvature
will be Ricci-flat. Thus it is not surprizing that the Kleinian
singularity can be blown up to obtain a family  of complete non-singular
self-dual metrics which are ALE (Asymptotically Locally Euclidean).
These Riemannian self-dual metrics, or gravitational instantons as they
are sometimes called, depend upon a finite number of moduli. In our
case, with signature $(2,2)$ the moduli space is, by contrast, infinite
dimensional. Another difference is that because the support of the
curvature is zero dimensional from the orbifold construction of the
self-dual instantons, the usual Regge calculus approach to distributional
metrics, where curvature has support only on two-dimensional sets,
does not apply. What is needed is a formalism in which the
quadratic functions of the Riemann tensor that gives rise to the Euler
density and the Hirzebruch density have support at the origin.

Let us return to our solution. It is easy to see that the metric admits
two covariantly constant spinors, since the curvature is self-dual. In
addition there is a further covariantly constant spinor of opposite
chirality. Let us call these spinors $o_A, \iota_A$ and
$\tilde o_{A'}$ respectively. It follows immediately from this
that the metric must admit two covariantly constant null vectors given
by $o_A\tilde o_{A'}$ and $\iota_A\tilde o_{A'}$ These vectors
automatically obey Killings equation, and span the covariantly constant
null 2-surface with 2-form $\tilde o_{A'}\tilde o_{B'}o_Ao_B$
which is self-dual. This corresponds to our privileged family of
$\alpha$-planes.

It is tempting to identify these spacetimes with the physical spacetime
in which an $N=2$ superstring moves, and the cosmic string with the
2-surface itself. The existence of the covariantly constant spinors is
promising because it shows that the spacetime can admit unbroken
supersymmetry as one might expect.

\end